\documentclass{aa} 

\usepackage{ulem}
\usepackage{graphicx}
\usepackage[varg]{txfonts}
\usepackage{color}

\usepackage{natbib}
\bibpunct{(}{)}{;}{a}{}{,} 

\newcommand{\micron}{\mathrm{\mu m}}

\begin{document} 


   \title{A likely detection of a local interplanetary dust cloud passing near the Earth in the AKARI mid-infrared all-sky map}
   \titlerunning{A likely detection of a local cloud in the AKARI MIR all-sky maps}

   \author{D.~Ishihara\inst{1}\and T.~Kondo\inst{1}\and H.~Kaneda\inst{1}\and T.~Suzuki\inst{1}\and 
K.~Nakamichi\inst{1}\and S.~Takaba\inst{1}\and H.~Kobayashi\inst{1}\and S.~Masuda\inst{2}\and T.~Ootsubo\inst{3}\and J.~Pyo\inst{4}\and T.~Onaka\inst{5}
          }

   \institute{Graduate School of Science, Nagoya University, Chikusa-ku, Nagoya 464-8602, Japan\\
              \email{ishihara@u.phys.nagoya-u.ac.jp}
         \and
             Institute for Space-Earth Environmental Research, Nagoya University, Chikusa-ku, Nagoya 464-8601, Japan
         \and
             Graduate School of Arts and Sciences, The University of Tokyo, Meguro-ku, Tokyo 153-8902, Japan
         \and
                     Korea Astronomy and Space Science Institute, Daejeon 305-348, Republic of Korea
                 \and
                     Graduate School of Science, The University of Tokyo, Bunkyo-ku, Tokyo 113-0033, Japan
             }

   \date{Received / Accepted}

 
  \abstract
   {We are creating the AKARI mid-infrared all-sky diffuse maps. 
Through a foreground removal of the zodiacal emission,
we serendipitously detected a bright residual component whose angular size is 
about $50\degr\times 20\degr$ at a wavelength of 9 $\micron$.}
   {We investigate the origin and the physical properties of the residual component.}
   {We measured the surface brightness of the residual component in the AKARI mid-infrared all-sky maps.}
   {The residual component was significantly detected only in 2007 January, even though the same region was observed in 2006 July and 2007 July, which   shows that it is not due to the Galactic emission. We suggest that this may be a small cloud passing near the Earth.
By comparing the observed intensity ratio of $I_{9\,\mu m}/I_{18\,\mu m}$
with the expected intensity ratio assuming thermal equilibrium of dust grains at 1~AU 
for various dust compositions and sizes,
we find that dust grains in the moving cloud are likely to be much smaller than typical grains that produce the bulk of the zodiacal light.
}
   {Considering the observed date and position, it is likely that it originates in the solar coronal mass ejection (CME) which took place on 2007 January 25. 
}


   \maketitle
%

\section{Introduction}
\label{intro}

AKARI \citep{murakami07}, the Japanese infrared (IR) astronomical satellite, carried out all-sky surveys at wavelengths of 9, 18, 65, 90, 140, and 160 $\micron$ during the period from 2006 May 6 to 2007 August 28. The mid-IR (9 and 18 $\micron$) data obtained by the Infrared Camera (IRC; \citealt{onaka07}), one of the focal-plane instruments aboard the AKARI satellite, are currently being processed for  public release of the all-sky diffuse maps, while the far-IR (65, 90, 140, and 160 $\micron$) all-sky diffuse maps obtained by the Far-Infrared Surveyor (FIS; \citealt{kawada07}), the other focal-plane instrument, have been already released to the public \citep{doi15}. 

\begin{figure*}
\centering
        \includegraphics[width=17cm]{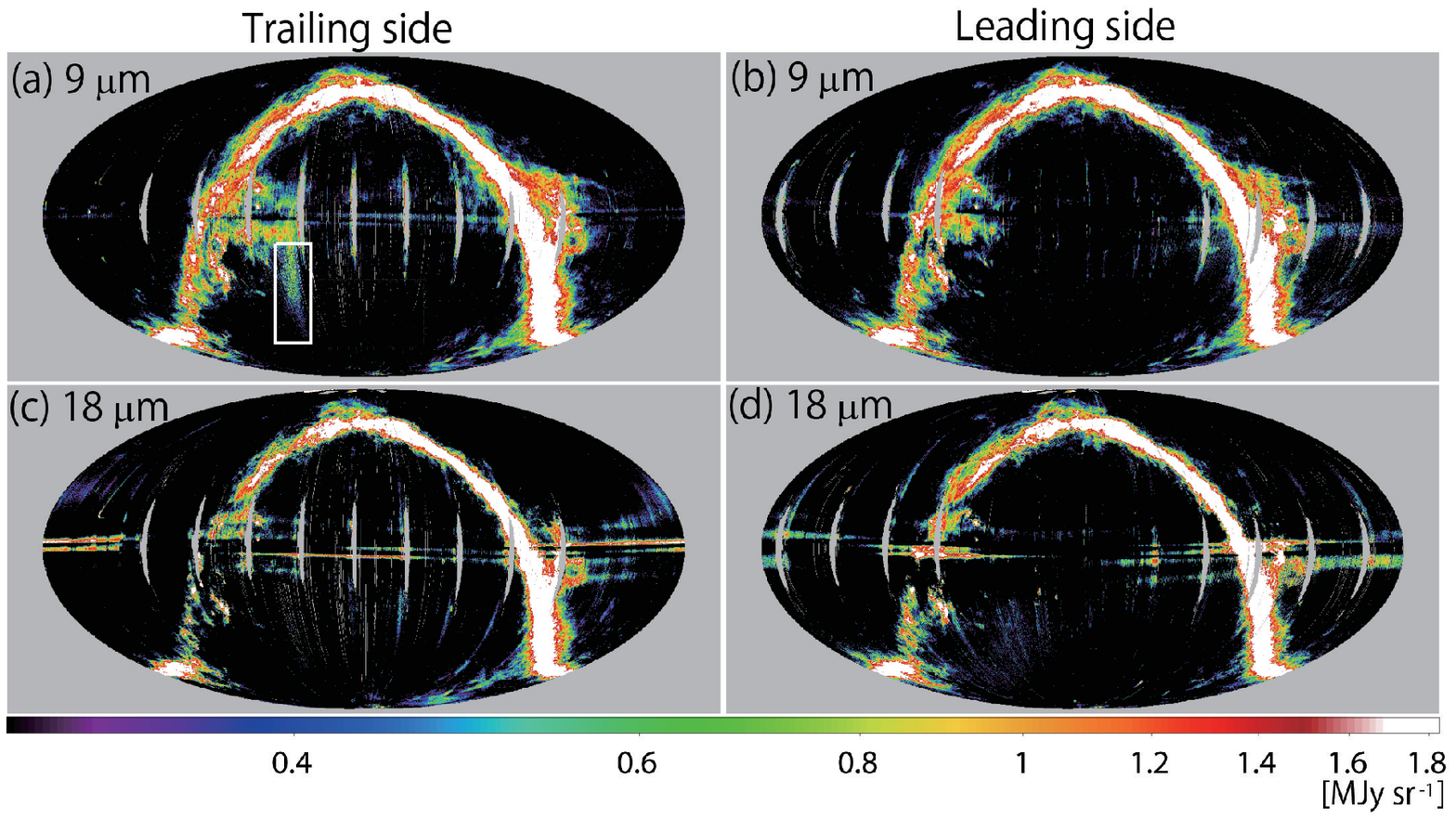}
                \caption{AKARI mid-IR all-sky diffuse maps in the ecliptic coordinates after subtraction of the zodiacal emission with the \citet{kondo16} model: a) at 9 $\micron$ on the trailing side, b) at 9 $\micron$ on the leading side, c) at 18 $\micron$ on the trailing side, and d) at 18 $\micron$ on the leading side. The colour scales are the same in all four panels. The solid box in panel (a) indicates the position of the extended bright residual component. This area is $18\degr\times 50\degr$ with the centre at $(\lambda ,\beta )=(45\degr ,-36\degr )$. }
                \label{fig:allsky}
\end{figure*}

The subtraction of the zodiacal emission, the thermal IR emission from interplanetary dust (IPD) in our solar system, 
is one of the most important data reduction procedures for galactic and extragalactic sciences. Through this process of the mid-IR all-sky data,
we have found a bright residual component in the 9 $\micron$ band, which is extended around 
$(\lambda ,\beta) =(45\degr ,-30\degr)$, as shown in Fig. \ref{fig:allsky}a,
where $\lambda$ and $\beta$ are the ecliptic longitude and latitude, respectively.
AKARI observed this region three times, in 2006 July, 2007 January, and 2007 July,
but the residual component was observed only in 2007 January. Therefore, this is likely to be a temporal interplanetary component in our solar system. 
The distribution of the interplanetary dust (IPD) is not generally considered to vary on such a short timescale as $\sim 1$ year because the main contribution of the IPD is dust grains migrating from the outer regions of our solar system by the Poynting-Robertson effect on a timescale of about $10^4$ years
\citep{burns79}.
On the other hand, \citet{kondo16} find that the distribution of the mean motion resonance component, trapped by the Earth into resonant orbits near 1~AU, has changed during the  16 years between the COBE and the AKARI epoch;
this change in distribution may be caused by the dust supply with comets near the Earth orbit.
%
%
This suggests that local structures of the IPD cloud can change on timescales shorter than a few decades. In this paper we aim to reveal the origin and the physical properties of the temporal component which appeared only in 2007 January.


\section{Observations and data reduction}
\label{obs}

AKARI was launched on 2006 February 21 (UT) and brought into a Sun-synchronous polar orbit at an altitude of 700 km. AKARI revolved around the Earth and scanned the sky along the circle of the solar elongation at approximately $90\degr$ in the all-sky survey. The orbit rotated around the axis of the Earth in one year, and hence the satellite covered the whole sky in half a year. The all-sky survey observations were carried out for about one year and three months with the telescope cooled at 6 K by super-liquid helium and mechanical coolers \citep{kaneda05,kaneda07}. Since the solar elongation was fixed at $90\degr$, AKARI observed the trailing and the leading direction of the Earth orbit alternately for every scan.

The mid-IR all-sky survey was conducted with the two photometric broad-band filters centred at 9 and 18 $\micron$ of the MIR-S and MIR-L channels, which have $256\times 256$ pixels with the pixel scales of $2\farcs 34\times 2\farcs 34$ and $2\farcs 51\times 2\farcs 39$, respectively \citep{onaka07,ishihara10}. The full widths at half maxima of the point spread functions for the 9 and 18 $\micron$ bands are $5\farcs 5$ and $5\farcs7$, respectively \citep{ishihara10}. The spectral response curves of these bands are shown in Fig. 1 in \citet{ishihara10}.

In order to produce the mid-IR all-sky diffuse maps, the data processing was carried out in addition to the original processes for the point source detection and measurement described in \citet{ishihara10}. The correction for the ionizing radiation effects induced by high-energy cosmic rays and the subtraction of the zodiacal emission are described in detail in \citet{mouri11} and \citet{kondo16}, respectively. The other processes such as corrections for the reset anomaly and non-linearity of the detector and that for scattered light from the Moon will be described in the official document of the AKARI mid-IR all-sky diffuse maps being prepared for the public data release (Ishihara et al. in prep.).

\begin{figure}
  \resizebox{\hsize}{!}{\includegraphics{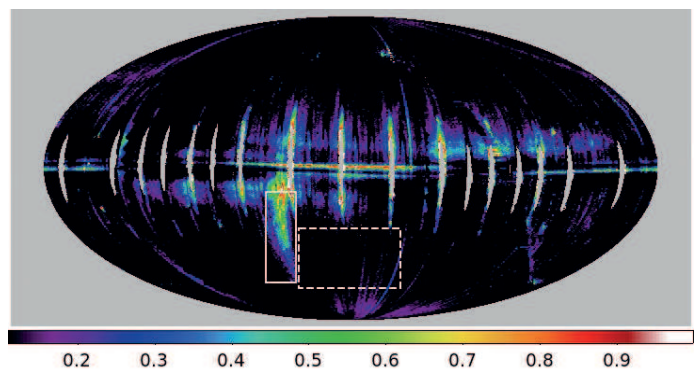}}
  \caption{Difference between the trailing-side map and the leading-side map 
in the 9~$\micron$ band. 
The position and the size of the solid box is the same as that in Fig.~\ref{fig:allsky}a.
The dashed box indicates the area of $54\degr\times 34\degr$ 
with the centre at $(\lambda ,\beta )=(354\degr ,-47\degr )$, 
where the brightness fluctuations are measured.}
  \label{fig:diff}
\end{figure}

Figure \ref{fig:allsky} shows the AKARI mid-IR all-sky diffuse maps after the data processing. 
The regions affected by scattered light from the Moon are masked along the ecliptic latitude with the length of $\sim 30\degr$ near the ecliptic plane. 
The position of the residual component in the 9 $\micron$ band which appeared in 2007 January is indicated by the solid box in Fig. \ref{fig:allsky}a, 
which is the trailing-side map at 9 $\micron$. 
%
%
Figure \ref{fig:diff} shows the difference map between 
the trailing and the leading sides at 9 $\micron$, which clearly exhibits the extended residual component.
%
The mean brightness of the component is measured 
in the solid box
of $18\degr\times 50\degr$, while the error is estimated from the brightness 
fluctuation from the adjacent dashed box in  Fig.~\ref{fig:diff}, 
which is selected to avoid the galactic and the ecliptic plane regions. 
We also measured the mean brightnesses in the same area in Figs. \ref{fig:allsky}a to d.


\section{Results}
\label{results}

Table \ref{table:1} shows the mean brightnesses in the rectangular area
indicated by the solid box in Fig.~\ref{fig:allsky}. 
As shown in this table, the extended bright residual component 
is detected with significance higher than $5\sigma$ at 9 $\micron$ 
in the trailing-side map, while it is not detected with $2\sigma$ significance in the other maps. 
This suggests that the component appeared only on the trailing side, 
and therefore it cannot be considered  the galactic component.  
The regions where the surface brightness is higher than 0.22 MJy~sr$^{-1}$ ($5\sigma$ level) 
in Fig.~\ref{fig:diff} extend over $50\degr$ along the ecliptic latitude, 
which is not likely to be either the residual of the scattered light 
from the Moon near the ecliptic plane or the residual of the zodiacal emission along the ecliptic plane. 
Moreover this component was detected for about 15 consecutive days. 
Therefore, it cannot be explained by any instrumental artefacts, 
and this is probably a temporal component in our solar system, 
such as a small cloud passing near the Earth which was serendipitously detected by the AKARI all-sky survey.
%

\begin{table*}
\caption{Mean brightness in the rectangular area indicated by the solid box in Fig. \ref{fig:allsky}}
\label{table:1}
\centering
\begin{tabular}{c l l c c}
\hline\hline
 & Wavelength & Direction & Mean brightness (MJy~sr$^{-1}$) & Error (MJy~sr$^{-1}$) \\
\hline
(a) & 9 $\micron$ & Trailing & 0.25 (0.46\tablefootmark{a}) & 0.05 \\
(b) & 9 $\micron$ & Leading & 0.09 & 0.05 \\
(c) & 18 $\micron$ & Trailing & $-0.06$ & 0.07 \\
(d) & 18 $\micron$ & Leading & 0.09 & 0.11 \\
\hline
\end{tabular}
\tablefoot{
\tablefoottext{a}{The mean brightness
measured from the regions where the surface brightness is higher than 0.22 MJy~sr$^{-1}$ (5$\sigma$ level).}
}
\end{table*}


\section{Discussion}
\label{discussion}

\subsection{Origin of the likely local cloud}
\label{origin}

\begin{figure*}
\centering
\includegraphics[width=\textwidth]{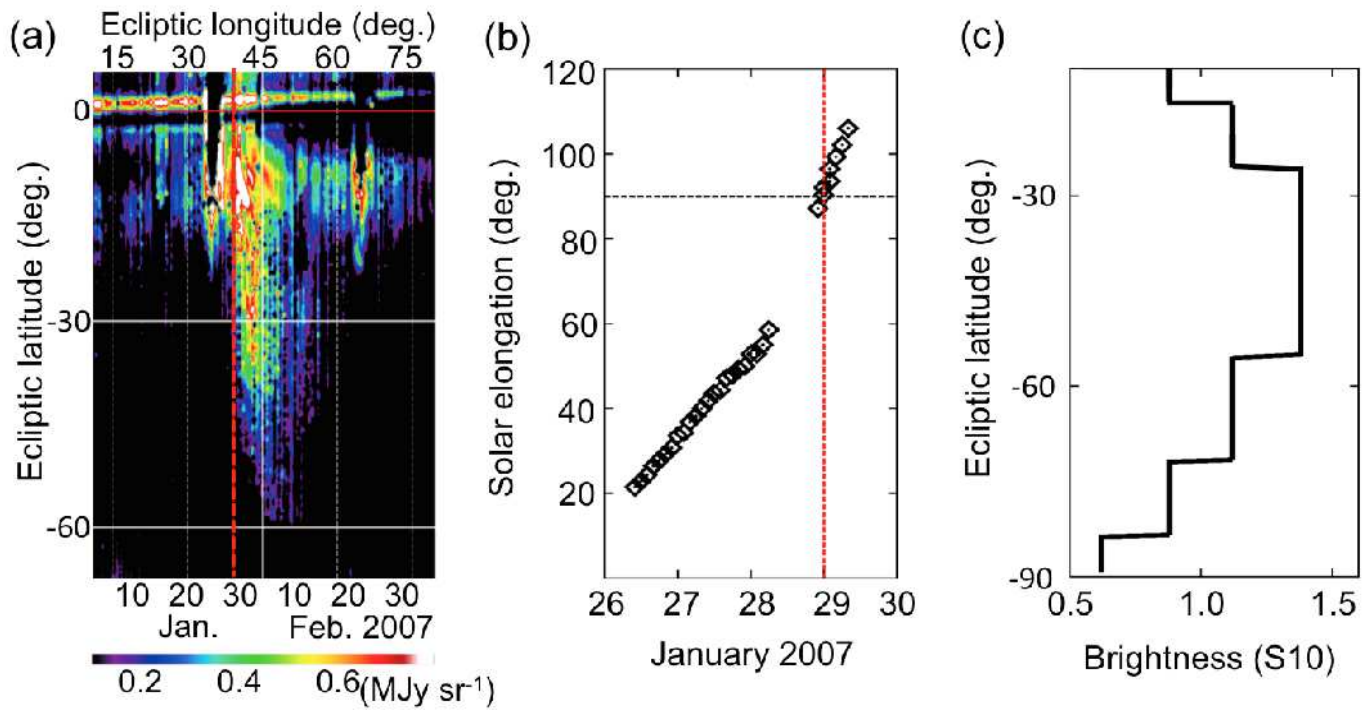}
\caption{
a) Enlarged view of the difference between the trailing-side map and the leading-side map
in the 9 $\micron$ band around the extended bright residual component.
The corresponding observation date is shown in the lower horizontal axis.
The direction of the horizontal axis is opposite to that in Fig.~\ref{fig:diff}. 
The likely local cloud is recognized for about 15 days from 2007 January 29 as indicated by the red line.
b) Solar elongation of the CME front as a function of date, obtained by observations with
the STEREO/HI2 and the Coriolis/SMEI \citep{webb09}. 
The red line indicates the same date as that in panel a,
while the black dotted line corresponds to the solar elongation of $90\degr$.
c)  CME brightness profile taken by the SMEI 
in the south region on the trailing side along the solar elongation of $90\degr$ on 2007 January 29.}
        \label{fig:day}
\end{figure*}

The existence of the temporal component, which is likely to be a local cloud passing near the Earth, is recognized
for about 15 days starting on 2007 January 29, as shown in Fig. \ref{fig:day}a. 
One possibility is a temporal change of the trailing cloud
following the Earth, such as that reported in \citet{kondo16}.
As shown later, however,
the estimated dust sizes are significantly smaller than
those expected in this case and therefore this case is unlikely.
Another possibility is either a relatively recent collision of asteroids
or the seeing material in a comet's orbit.
The former possibility is, however, unlikely 
since such a collision should have been reported by the Near Earth Object Program.\footnote{\url{http://neo.jpl.nasa.gov/}}
The latter is not likely either
since this transient cloud was observed for about 15 consecutive days.

Yet another possibility is that it originates in the solar coronal mass ejections (CMEs). 
 The CME event occurred on 2007 January 25 (e.g. \citealt{grigoryeva09,webb09}) and the CME front was observed at the solar elongation of $90\degr$ on 2007 January 29 by the Solar Mass Ejection Imager (SMEI) aboard the Coriolis satellite, as shown in Fig. \ref{fig:day}b. The interplanetary CME brightness distribution taken with SMEI can be seen on the SMEI website.\footnote{\url{http://smei.ucsd.edu/new_smei/index.html}} The CME distribution on 2007 January 29 shows that the brightness is relatively high in the south region on the trailing side of the Earth, which roughly corresponds to the region where the likely local cloud was detected in the AKARI map. 
In Fig.~\ref{fig:day}c
we show the CME brightness profile in the south region on the trailing side along the solar elongation
of 90$^\circ$.

If the local cloud is of a CME origin, the cloud is
likely to consist of nanometer-sized dust grains; 
 the STEREO/WAVES data showed that the probability of detecting the flux of nanometer-sized dust grains becomes relatively high when the CMEs occur \citep{chat15}. 
The nanometer-sized dust grains whose velocities are about 100--1000~km~s$^{-1}$ have been detected 
at around 1~AU by the International Space Station \citep{carpenter07} and the STEREO spacecraft \citep{meyer09}. 
\citet{czechowski10} has carried out simulations of grain dynamics to show that nanometer-sized dust grains in the inner regions of our solar system are accelerated to a solar wind speed through interaction with interplanetary magnetic field because of high ratios of the charge to mass of the nanometer-sized dust grains. \citet{chat15} suggested -- using the STEREO/WAVE data --
 that the nanometer-sized dust grains, which have already been accelerated by the above mechanism, are transferred to around 1~AU with the CMEs.

However, the scale of the CME event which occurred on 2007 January 25 was not so large and hence not rare.
According to the SOHO LASCO CME Catalog\footnote{\url{http://cdaw.gsfc.nasa.gov/CME_list/}}, 
a CME event of this class occurred at least four times during the AKARI observation period.
Nevertheless, we did not observe corresponding significant increases 
in the brightness for the other CME events.
This suggests that only the arrival of CMEs at the Earth  does not necessarily explain 
the transfer of nanometer-sized dust grains to near the Earth. 
Indeed, \citet{chat15} showed that there is no time correlation between the frequency of CME events and 
the cumulative flux of the nanometer-sized dust grains observed by the STEREO/WAVES. 
The amount of the dust grains transferred by CMEs should depends on
the total amount integrated along the migration paths of CMEs.
A geometrical effect may also contribute to 
the difference in the observed brightness 
(e.g. whether or not a sheet-like material is viewed edge-on);
the elongated appearance of the residual component seen in Fig.~\ref{fig:allsky}
may indicate a non-spherical geometry.
%

\subsection{Physical properties of the likely local cloud}
\label{properties}

\begin{figure}
  \resizebox{\hsize}{!}{\includegraphics{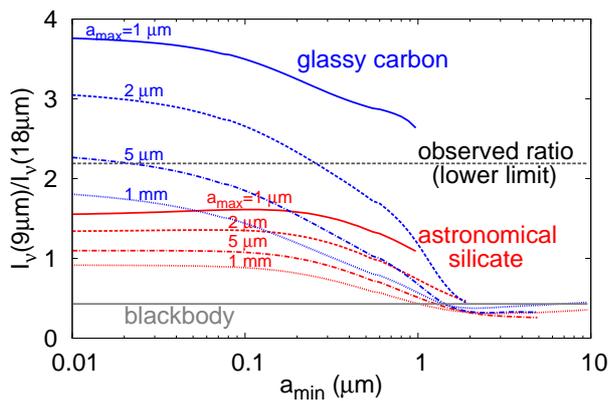}}
  \caption{Observed intensity ratio (lower limit) of
$I_{\rm \nu}$(9~$\mu$m) / $I_{\rm \nu}$(18~$\mu$m) (black dashed line) is
compared with expected intensity ratios as a function of $a_{\rm min}$
for different $a_{\rm max}$ and dust compositions.
The red, blue, and gray curves indicate 
astronomical silicate, glassy carbon, and bulk dust emitting blackbody radiation, respectively.
For astronomical silicate and glassy carbon, 
the solid, dashed, dash-dotted, and dotted curves indicate
$a_{\rm max}$ of 1\,$\mu$m, 2\,$\mu$m, 5\,$\mu$m, and 1\,mm, respectively.
}
  \label{fig:dust}
\end{figure}

We put constraints on the size distribution and composition of dust grains 
in the likely local cloud from the observed intensities.
In the thermal equilibrium, the temperature of a dust grain of size $a$, $T(a)$, is calculated from
\begin{equation}
\pi a^2 \frac{R_\odot^2}{R^2} 
\int{Q_{\rm abs}(a, \lambda) B_\lambda(\lambda, T_\odot)} d\lambda
= 4\pi a^2 \int{Q_{\rm abs}(a, \lambda) B_\lambda(\lambda, T(a))}d\lambda,
\label{eq:2}
\end{equation}
where
$Q_{\rm abs}(a,\lambda)$ is absorption efficiency for a grain of size $a$,
$B_\lambda(\lambda, T(a))$ is the Planck function,
$R$ is the distance from the Sun ($\geq$1~AU), and 
$R_\odot$ and $T_\odot$ are the radius and temperature of the Sun.
Using $T(a)$ for each grain size,
the surface brightness of the emission from the dust cloud at a wavelength of $\lambda$ 
is indicated as
\begin{equation}
I_\lambda(\lambda) = {\int_{a_{\rm min}}^{a_{\rm max}} \pi a^2 Q_{\rm abs}(\lambda, a) B_\lambda(\lambda, T(a)) N(a) da},
\label{eq:1}
\end{equation}
where $N(a) da$ is the number column density of the grains with  sizes between $a$ and $a+da$,
and $a_{\rm min}$ and $a_{\rm max}$ are the minimum and maximum sizes of the grains, respectively.
For $Q_{\rm abs}$, we consider the following dust compositions:
astronomical silicate \citep{draine2003}, 
glassy carbon \citep{hanner87}, and bulk dust emitting blackbody radiation.
We assume that
the size distribution of the grains is given by $N(a) = N_0 \cdot a^{-3.5}$ 
as seen in the solar system \citep{grun85,reach88}.
The values of $a_{\rm max}$ is set to be 1~$\mu$m, 2~$\mu$m, 5~$\mu$m, and 1~mm
because the result changes drastically around $a_{\rm max}\approx$ a few~$\mu$m.
For each of the compositions and $a_{\rm max}$, we investigate
$a_{\rm min}$ for the range between 10\,nm and 10\,$\mu$m.
From equation \ref{eq:1}, the observed intensity ratio
of 9\,$\mu$m over 18\,$\mu$m 
is expressed as a function of $a_{\rm min}$ and $a_{\rm max}$ for each composition as
\begin{equation}
{\rm Ratio} (a_{\rm min},a_{\rm max}) = \frac
{\int_{a_{\rm min}}^{a_{\rm max}} a^2 Q_{\rm abs}(9\micron, a) B_\lambda(9\micron, T(a)) N(a) da}
{\int_{a_{\rm min}}^{a_{\rm max}} a^2 Q_{\rm abs}(18\micron, a) B_\lambda(18\micron, T(a)) N(a) da}.
\label{eq:3}
\end{equation}
The 9 $\micron$ mean brightness of the likely local cloud is calculated to be 0.46\,MJy~sr$^{-1}$ 
by using the intensities of the pixels with significant ($>5\sigma$) detection,
while the $3\sigma$ upper limits of the 18 $\micron$ brightness is calculated to be 0.21\,MJy~sr$^{-1}$.
By comparing the observed intensity ratio with
the expected intensity ratios based on equation \ref{eq:3}, 
we put constraints on $a_{\rm min}$ and $a_{\rm max}$ values for each composition.
Figure~\ref{fig:dust} summarizes the results for $R=1$~AU.
From the figure, we find that
the observed intensity ratio cannot be explained
by any $a_{\rm min}$ and $a_{\rm max}$ values for astronomical silicate and blackbody dust.
%
For glassy carbon, the observed intensity ratio is explained if
$a_\mathrm{max} = 1$\,$\mu$m without constraints on $a_\mathrm{min}$.
For $a_{\rm max}$=2 and 5~$\micron$, $a_{\rm min}$ is expected to be $<300$~nm and $<10$~nm, respectively.
For $a_\mathrm{max} > 6$\,$\mu$m, the observed colour cannot be explained even by carbon dust.
If $R$ is larger than 1~AU, the equilibrium temperatures decrease,
thus the grain sizes to explain the observed ratio become even smaller.
The likely size distributions 
are weighted toward smaller sizes than that of typical IPD grains \citep[$>$10\,$\mu$m;][]{grun85,reach88}.

Then, we calculated the mass density for the observed $I_\lambda$(9\,$\mu$m),
under the constraints on $a_{\rm min}$, $a_{\rm max}$, and composition obtained above.
The mass column density of the cloud, $\Sigma_{\rm dust}$, is given by
\begin{equation}
\Sigma_{\rm dust} =
{\int_{a_{\rm min}}^{a_{\rm max}} \left(\frac{4}{3}\pi a^3 \rho\right) N_0 a^{-3.5} da} ,
\label{eq:4}
\end{equation}
assuming spherical grains,
where $\rho$ is the specific mass of glassy carbon (2.2 g ${\rm cm^{-3}}$). The value of 
$N_0$ is calculated from equation 2 for $\lambda$=9~$\micron$.
From equation \ref{eq:4}, we obtain $\Sigma_{\rm dust} =$ 1--3$\times10^{-14}$\,g\,cm$^{-2}$.
Assuming that the line-of-sight depth of the likely local cloud is 0.1~AU, 
the mass density range is estimated as 1--2$\times10^{-26}$\,g\,cm$^{-3}$,
which is 3 orders of magnitude smaller than
that of the interplanetary dust cloud at 1~AU from the Sun (e.g. \citealt{rowan13}).
The overall results on $a_{\rm min}$ and $\Sigma_{\rm dust}$ 
indicate that the origin of the local cloud is different from those of normal IPD clouds,
and are consistent with the picture that the local cloud is of a CME origin.

\section{Summary}

A widely extended bright component whose angular size is about
$50\degr\times 20\degr$ was found in the AKARI 9 $\micron$ all-sky
diffuse map. This component was detected when the satellite was on the
trailing side of the Earth's orbit, but not when the satellite was on the
leading side while the telescope observed the same celestial
direction. Therefore, it is not due to the galactic emission. The
morphology indicates that it does not originate from the residual of
subtraction of either the zodiacal emission or the scattered light from
the Moon. The large structure, which continued to be detected for about
15 days, cannot be explained by any instrumental artefacts. Therefore,
this is likely to be a small cloud passing near the Earth which was
serendipitously detected by the AKARI all-sky survey. Considering the
observed date and position of the cloud, it may be related to the solar
CME which took place on 2007 January 25. Since  previous studies
have suggested that CMEs contain nanometer-sized dust grains, the 9 $\micron$
emission detected by the AKARI all-sky survey is likely to be attributed
to such dust grains. 
By comparing the observed colour temperature
with the equilibrium temperature estimated at 1~AU,
dust grains in the moving cloud are likely to be
much smaller than typical IPD grains.
If the line-of-sight depth of the cloud is 0.1~AU, 
the dust mass density is estimated 
to be much smaller than that of a typical IPD cloud around 1~AU.
These results are therefore
consistent with the picture that the moving cloud is of a CME origin.


\begin{acknowledgements}

We thank all the members of the AKARI project. AKARI is JAXA project with the participation of ESA. 
The CME catalog is generated and maintained at CDAW Data Center 
by NASA and The Catholic University of America in cooperation with the Naval Research Laboratory. 
SOHO is a project of international cooperation between ESA and NASA.
This research is financially supported by Grants-in-Aid for 
Young Scientists (A) No.~26707008, and Scientific Research (C) No.~25400220.
T.K. is financially supported by Grant-in-Aid for JSPS Fellows No.~25002536, and the Nagoya University
 Program for Leading Graduate Schools, ``Leadership Development Program
for Space Exploration and Research,'' from MEXT.
\end{acknowledgements}

\bibliographystyle{aa}
\bibliography{refer}

\begin{thebibliography}{18}
\expandafter\ifx\csname natexlab\endcsname\relax\def\natexlab#1{#1}\fi

\bibitem[{Burns {et~al.}(1979)}]{burns79}
Burns, J. A., Lamy, P. L., \& Soter, S. 1979, \icarus, 40, 1

\bibitem[{Carpenter {et~al.}(2007)Carpenter, Stevenson, Fraser,
  {et~al.}}]{carpenter07}
Carpenter, J.~D., Stevenson, T.~J., Fraser, G.~W., {et~al.} 2007, Journal of
  Geophysical Research (Planets), 112, E08008

\bibitem[{Czechowski \& Mann(2010)}]{czechowski10}
Czechowski, A. \& Mann, I. 2010, \apj, 714, 89

\bibitem[{Doi {et~al.}(2015)Doi, Takita, Ootsubo, {et~al.}}]{doi15}
Doi, Y., Takita, S., Ootsubo, T., {et~al.} 2015, \pasj, 67, 50

\bibitem[{Draine(2003)}]{draine2003}
Draine, B.~T. 2003, \araa, 41, 241

\bibitem[{Grigoryeva {et~al.}(2009)Grigoryeva, Borovik, Livshits,
  {et~al.}}]{grigoryeva09}
Grigoryeva, I.~Y., Borovik, V.~N., Livshits, M.~A., {et~al.} 2009, \solphys,
  260, 157

\bibitem[{Gr\"{u}n {et~al.}(1985)Grun, Zook, Fechtig, {et~al.}}]{grun85}
Gr\"{u}n, E., Zook, H. A., Fechtig, H., and Giese, R. H., 1985, \icarus, 62, 244

\bibitem[{Ishihara {et~al.}(2010)Ishihara, Onaka, Kataza,
  {et~al.}}]{ishihara10}
Ishihara, D., Onaka, T., Kataza, H., {et~al.} 2010, \aap, 514, A1

\bibitem[{Hanner (1987) Hanner}]{hanner87}
Hanner, M. 1987, Grain optical properties, in Infrared Observations of Comets Halley \& Wilson and Properties of the Grains, edited by M. Hanner, pp. 22-49, NASA Conference, Washington.




\bibitem[{Kaneda {et~al.}(2007)Kaneda, Kim, Onaka, {et~al.}}]{kaneda07}
Kaneda, H., Kim, W., Onaka, T., {et~al.} 2007, \pasj, 59, 423

\bibitem[{Kaneda {et~al.}(2005)Kaneda, Onaka, Nakagawa, {et~al.}}]{kaneda05}
Kaneda, H., Onaka, T., Nakagawa, T., {et~al.} 2005, \ao, 44, 6823

\bibitem[{Kawada {et~al.}(2007)Kawada, Baba, Barthel, {et~al.}}]{kawada07}
Kawada, M., Baba, H., Barthel, P.~D., {et~al.} 2007, \pasj, 59, 389


\bibitem[{Kondo {et~al.}(2016)Kondo, Ishihara, Kaneda, {et~al.}}]{kondo16}
Kondo, T., Ishihara, D., Kaneda, H., {et~al.} 2016, \aj, 151, 71

\bibitem[{Le~Chat {et~al.}(2015)Le~Chat, Issautier, Zaslavsky,
  {et~al.}}]{chat15}
Le~Chat, G., Issautier, K., Zaslavsky, A., {et~al.} 2015, \solphys, 290, 933


\bibitem[{Meyer-Vernet {et~al.}(2009)Meyer-Vernet, Maksimovic, Czechowski,
  {et~al.}}]{meyer09}
Meyer-Vernet, N., Maksimovic, M., Czechowski, A., {et~al.} 2009, \solphys, 256,
  463

\bibitem[{Mouri {et~al.}(2011)Mouri, Kaneda, Ishihara, {et~al.}}]{mouri11}
Mouri, A., Kaneda, H., Ishihara, D., {et~al.} 2011, \pasp, 123, 561

\bibitem[{Murakami {et~al.}(2007)Murakami, Baba, Barthel,
  {et~al.}}]{murakami07}
Murakami, H., Baba, H., Barthel, P., {et~al.} 2007, \pasj, 59, 369


\bibitem[{Onaka {et~al.}(2007)Onaka, Matsuhara, Wada, {et~al.}}]{onaka07}
Onaka, T., Matsuhara, H., Wada, T., {et~al.} 2007, \pasj, 59, 401


\bibitem[{Reach (1988)}]{reach88}
Reach, W. T., 1988, \apj, 335, 468

\bibitem[{Rowan-Robinson \& May(2013)}]{rowan13}
Rowan-Robinson, M. \& May, B. 2013, \mnras, 429, 2894

\bibitem[{Webb {et~al.}(2009)Webb, Howard, Fry, {et~al.}}]{webb09}
Webb, D.~F., Howard, T.~A., Fry, C.~D., {et~al.} 2009, \solphys, 256, 239

\end{thebibliography}
\end{document}